\documentclass{kapproc} 

\usepackage[T1]{fontenc}

\usepackage{procps} 

\usepackage[dvips]{graphicx}

\upperandlowercase

\kluwerbib 

\begin{document}


\articletitle[]{Starless Cores}



 \author{Mario Tafalla}
 \affil{Observatorio Astron\'omico Nacional, Alfonso XII 3, E-28014 
 Madrid, Spain}
 \email{m.tafalla@oan.es}

 \begin{abstract}
Dense low mass cores in nearby clouds like Taurus and Auriga are some 
of the simplest sites currently forming stars like our Sun.
Because of their simplicity and proximity, dense cores offer the 
clearest view of the different phases of star formation, 
in particular the conditions prior to the onset of gravitational collapse.
Thanks to the combined analysis of the emission 
from molecular lines and the emission/absorption from dust grains, 
the last several years have seen a very rapid progress in our 
understanding of the structure and chemical composition of starless cores.
Previous contradictions between molecular tracers
are now understood to arise from core chemical inhomogeneities, which are 
caused by the selective freeze out of molecules onto cold dust grains.
The analysis of the dust emission and absorption, in addition,  has
allowed us to derive accurate density profiles, and has made finally 
possible to carry out self consistent modeling of the 
internal structure of starless cores. 
In this paper I briefly review the evolution of core studies 
previous to the current golden age, and show how multi-tracer 
emission can now be modeled in a systematic manner. Finally
I show how we can start to reconstruct the early history of core
formation taking advantage of the chemical changes in the gas.

 \end{abstract}



\section{Introduction}

Dense starless cores in nearby clouds like
Taurus and Auriga represent the simplest star-forming sites.
They collapse and
produce individual stars (or binaries) in almost isolation,
with apparently little influence from the surrounding
cloud or from previous generations of stars. For this simplicity and
proximity,
starless cores constitute ideal places  
to elucidate the still mysterious process by which 
interstellar matter collapses and forms gravitationally-bound
self-luminous objects.

The simplicity of star-formation in cores, unfortunately, comes at 
the price of missing some elements
of interest for this meeting, like clustering or the formation
of high-mass stars. Several reviews in this book  
provide information on these topics, and the reader is
referred to them to complement the material presented here.
Despite the above limitations, however, protostars formed in starless cores 
display most phenomena that we associate with the birth of a typical star. 
Gravitational motions, binary and disk formation, and bipolar outflow 
ejection all occur in
young stellar objects (YSOs) recently formed (or in the process
of formation) in low mass cores, and most of these phenomena were in fact
first identified in dense core environments. Thus, if we are to understand 
the basic principles and time evolution of most star-formation physics, 
studying dense cores offers a most promising approach.

When we observe a starless core, we are seeing a system that very likely
will collapse to form a star (it is still an open question
whether all starless cores are truly pre stellar). For this
reason starless cores offer a snapshot of the initial conditions of 
star formation, and even of the first stages of the process if they 
have already started to collapse. Deriving these conditions from observations
is therefore of great interest, and more than two decades
of dense core studies attest the enormous effort made in this direction.
For a number of reasons discussed below, the last five years or
so have witnessed a very rapid progress in the study of dense cores,
and numerous results and papers have appeared in the literature. In this 
contribution
I review some of this recent work, with emphasis on the study of the
internal structure of dense cores. In the pedagogical spirit of the workshop, 
I start with a brief review of the field. I have chosen a historical
point of view (my own simplistic summary of the history) to 
emphasize what we have learned, what are some of the most relevant papers
in the field, and where should be aiming to go in the future.
Contributions in this book by Malcolm Walmsley and Paola Caselli
treat related topics from different points of view and are highly 
recommended.

\section{``Classical'' studies of starless cores}

The systematic study of low mass dense cores has a tradition of more than two
decades, and starts in earnest with the first searches based on the 
optical inspection of Palomar Survey plates by Myers et al. (1983)
(see Lee \& Myers 1999 for an update of this work). These searches
were followed by radio studies of their molecular emission, mostly
NH$_3$ (e.g., Myers \& Benson 1983), to determine the
basic gas physical properties. Subsequent
correlation of dense core positions and the location of IRAS point sources
by Beichman et al. (1986) confirmed that dense cores are the sites
of ongoing star formation. IRAS and NIR observations, in addition, 
gave rise to a division of cores into star-containing and
starless, depending on whether
they harbored or not a central luminous object.
It should be emphasized, however, that this
classical distinction is based on a given level of sensitivity
achieved more than a decade ago, and estimated by Myers et al. (1987) as
corresponding to a luminosity of 0.1 L$_\odot$ for the distance of 
Taurus.  The recent revolution in mid IR sensitivity brought by the
Spitzer Space Telescope has forced some corrections in the old classification
of cores, and moved some cores previously classified as starless to the
star-containing category (Young et al. 2004). The distinction between 
starless and star-containing cores, however, is still a fundamental
one, in the sense that it distinguishes between before and after the
formation of a central object.

The 1980s represent the golden years of mapping and analysis of dense cores
based on their molecular line emission,
what we could call the "classical period''. The main results from
this decade are summarized in the monumental
paper by Benson \& Myers (1989), whose 15th birthday closely coincides
with the celebration of this meeting in October 2004. Benson \& Myers's work 
is based on arcminute-resolution observations of a single tracer (NH$_3$),
and concentrates on the global or average properties of cores.
After observing and analyzing 
more than 100 dense cores (both starless and star-containing),
these authors derive, among other properties,
typical masses of a few M$_\odot$, sizes of
about 0.1 pc, densities of a few 10$^4$ cm$^{-3}$, temperatures of 
10 K, and subsonic turbulent motions. To recognize the importance
of these results one only needs to notice that these observational
parameters constituted part of the input for the so-called standard 
model of star formation,
which crystallized almost at the same time (Shu, Adams, \& Lizano 1987).

The above set of dense core properties was not free from contradictions,
as already discussed in the almost simultaneous paper by Zhou et al. (1989).
These authors compared observations of 
cores in two high density tracers, CS and NH$_3$, and found systematic 
discrepancies between them.  The most severe discrepancies 
were a mismatch
in the position of the emission peaks of the two molecules, 
a factor of 2 larger size of the CS maps, 
and a factor of 2 broader CS lines. Although different mechanisms were invoked 
at the time to explain these CS/NH$_3$ discrepancies, the lack of a detailed 
picture of the internal structure of dense cores left the problem unsolved.
As we will see below, only now, after more than a decade of work by a 
large number of researchers,
we can claim a reasonable understanding of the internal properties of 
starless cores and finally put an 
end to the old contradiction between different tracers.

\section{Starless cores studies in the 1990s}

The 1990s witnessed an enormous progress in the characterization of 
dense cores in general and starless cores in particular. This progress
resulted from a decade of improvements in the techniques to
trace the gas and dust components of dense cores, and in the parallel
development of sophisticated tools to analyze the data.
The classical method of molecular-line tracing 
greatly benefited from the factor of several increase in 
spatial resolution afforded by the new generation of millimeter
and submillimeter telescopes like the IRAM 30m and the JCMT. It also
benefited from the advent of multipixel receiver arrays, like 
QUARRY and 
SEQUOIA at FCRAO, which have allowed the systematic mapping of large
areas at moderate resolution. As a result, it is now possible to
map and resolve large numbers of dense cores in relatively short
amounts of time (e.g., Caselli et al. 2002).

More important than the incremental
increase in our ability to trace the molecular emission of cores
has been the development of new techniques to sample the dust component. 
This component can be traced by its millimeter and submillimeter thermal
emission, which is optically thin and most likely arises from dust of close to
constant emission properties. Pioneer work using mm and submm continuum 
observations has provided
the first realistic density profiles of starless cores, and has shown that
they present a characteristic flattening at their center (Ward-Thompson et al. 
1994, Andre et al.  1996). Such a flattening implies that the dust (and gas) 
density in the inner 5,000-10,000 AU of a core has a close-to-constant 
value which typically ranges between 10$^5$ to 10$^6$ cm$^{-3}$ 
(Ward-Thompson et al. 1999).

Additional progress in the tracing of the dust component comes from
the technique of NIR extinction measurement, which uses the reddening 
of background stars
behind a core or a cloud to estimate the dust column density (Lada et al.
1994). Using this technique, Alves et al. (2001) has
produced an extremely detailed view of the B68 globule, and has shown 
that its density profile is very close to that expected for a close-to-critical
isothermal (Bonnor-Ebert) sphere. A related technique, 
which uses the extinction
by a core of the mid-IR diffuse background, has been used by Bacmann et al.
(2000) to complement the mm and submm continuum emission measurements by
estimating density profiles and confirming the presence of central flattening.

When the above dust measurements and molecular line
observations were combined to carry out 
a unified analysis of starless cores,
it became clear that many molecules do not sample 
the central and densest part of the cores, but that they disappear 
from the gas phase most likely due to their freeze out onto
cold dust grains (Kuiper et al. 1996, Kramer et al. 1999, 
Caselli et al. 1999, Tafalla et al. 2002). This freeze out of molecules
onto grains is not only of chemical interest by itself
(see Bergin \& Langer 1997 for pioneer modeling), but 
has important consequences for dense core studies. Molecular emission
is the only tool available to study gas kinematics,
so the disappearance of some species at high densities 
imposes a significant limitation in this type of studies. 
As a positive aspect, the presence of selective freeze out of molecules in
the central gas of dense cores helps to explain the contradictions between
CS and NH$_3$ observations described by Zhou et al. (1989), because these two
molecules behave differently at high densities: CS freezes out easily
and NH$_3$ does not (see Tafalla et al. 2002 for further details). 

The 1990s was also a decade of intense studies of dense core kinematics,
spurred by the identification of infall motions in the 
star-containing core B335
(Zhou et al. 1993). Among starless cores, L1544 was the first one found to 
present evidence
for inward motions (Myers et al. 1996, Tafalla et al. 1998), and subsequent
searches by Lee et al. (1999) have identified several additional cases.
These initial infall searches were made using molecular species like CS and 
H$_2$CO, now known to
freeze out at densities typical of dense core centers, so they were 
only sensitive to motions in the outer
layers of the cores. More recent infall studies 
take molecular freeze out into consideration,
and can therefore penetrate deeper into the gas. 


\section{The internal structure of the L1498 and L1517B cores}

To show the level of detail currently possible in the analysis of the
internal structure of starless cores, this section 
summarizes the recent study of 
two cores in the Taurus-Auriga complex, L1498 and L1517B.
These two cores (Fig. 1) were selected for their close-to-round shape and 
relative isolation, in addition to being starless (no IRAS
or 2MASS sources), and they have been subject to observations in the 1.2mm
continuum and a number of molecular lines using the FCRAO 14m, IRAM 30m 
and Effelsberg 100m telescopes (Fig. 2).  
A complete account of is this work is presented in Tafalla et al.
(2004; 2005, in preparation).


\begin{figure}[h]
\begin{center}
\resizebox{10cm}{!}{\includegraphics{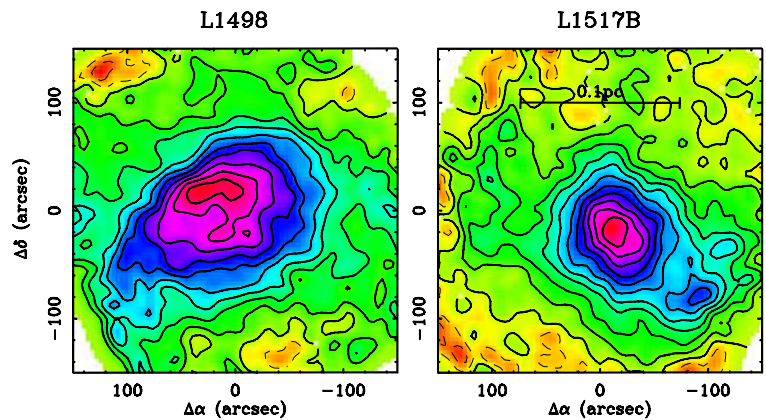}}
\end{center}
\caption{\protect\inx{1.2mm} continuum maps of L1498 and L1517B.
Note their central concentration and close-to-round geometries.
First contour and contour spacing are 2 mJy/$11''$-beam.}
\end{figure}


\begin{figure}[h]
\resizebox{\hsize}{!}{\includegraphics[angle=270,width=10cm]{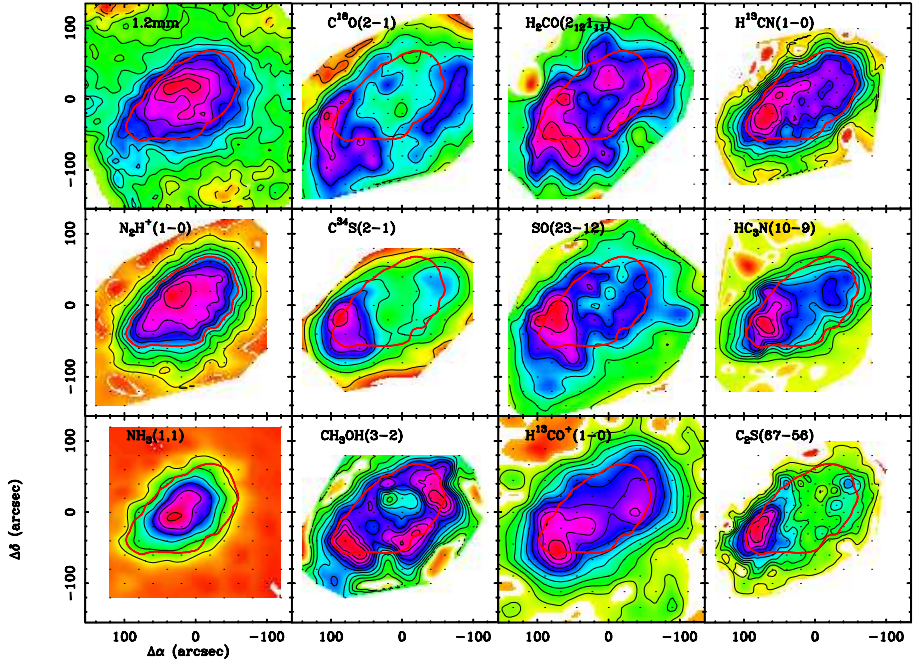}}
\caption{\protect\inx{Sample} of molecular emission maps of the L1498
core. Only the maps in the first column show centrally peaked emission,
while the rest present a dip indicative of molecular depletion. For easier
inter comparison, the half maximum contour of the N$_2$H$^+$(1--0) emission
has been superposed to each map (red line in the color version of the figure).}
\end{figure}

Characterizing the internal structure of a dense core implies deriving 
both its physical parameters (density, temperature, turbulence) and
its chemical properties (molecular abundances) as a function of position.
Given that we only see the cores projected on the plane of the sky,
we need to make an assumption on their third (line-of-sight) 
dimension, and for the close-to-round 
targets of our study, we will assume spherical symmetry.
Once we have fixed the geometry, we can use the optically thin 1.2mm continuum
emission to derive a density profile, and the observations of appropriate
molecular lines to derive the gas temperature, turbulence, and 
systemic motions in addition to the abundance profiles of each species.
Because of the gradient in gas excitation 
caused by the density 
gradient, a non-LTE radiative transfer solution is necessary
to model the molecular emission. For this work, we use the Monte Carlo 
radiative transfer code of Bernes (1979) updated with the most recent
molecular parameters and collision rates. This model is spherically symmetric, 
so we compare its results with radial averages of the emission,
and fit simultaneously the radial profile of integrated intensity and 
the central spectrum for
each line. The following subsections show the
analysis step by step and the conclusions derived from it.

\subsection{Density}

The first step in our analysis consists of deriving the density profile
of each core
based on its 1.2mm continuum emission. Unfortunately, this step is the
most uncertain one, due to our incomplete knowledge of
the physical properties of the emitting dust. In the optically thin case
(a good approximation at 1.2mm), the dust emission depends on the product of
the dust emissivity and the Planck function at the dust temperature, and 
both emissivity and dust temperature are not well constrained.
For this work, we assume that the two parameters have constant values of
0.005 cm g$^{-1}$ and 10 K, respectively. Our choice of dust temperature is
based on our estimate of the gas temperature using NH$_3$ and presented
in the next section, while our choice of emissivity follows the standard
value in the literature (e.g., Andre et al. 1996). If grain growth via
coagulation is important toward the core center, the true emissivity can
increase there, and be larger than our assumed value
by a factor of about two (Ossenkopf \& Henning 1994). The dust temperature,
on the other hand, may decrease toward the center due to the extinction 
of the interstellar radiation field (the dominant heating mechanism for 
dust in core environments), so it may reach a lower value of around 8 K
for our cores (Evans et al. 2001, Zucconi et al. 2001, Galli et al. 2002).
Thus, if for example we use
an emissivity of 0.009 cm g$^{-1}$, as recommended by Ossenkopf \& 
Henning (1994) (their OH5 value), and a dust temperature of 8.7 K,
as derived by Galli et al. (2002) for a core similar to L1517B,
we will have to decrease our densities by a factor 0.7.
Our chosen values, therefore represent a simplification of the complex
dust physics, but their product seems more immune to 
this physics that each of the individual components. 

\begin{figure}[h]
\begin{center}
\resizebox{12cm}{!}{\includegraphics{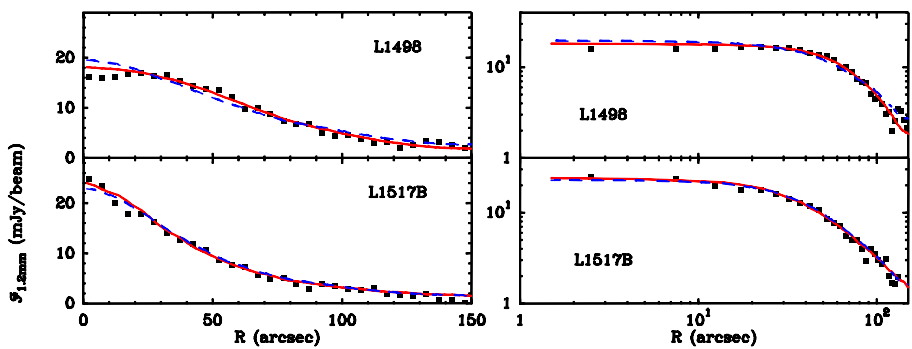}}
\end{center}
\caption{\protect\inx{Radial} profiles of 1.2mm continuum emission in
linear-linear and log-log scales (left and right, respectively). The
squares represent bolometer observations, the solid lines are the 
predictions from an analytic model, and the dashed line are the predictions
from an isothermal model (see text).}
\end{figure}

Once we have chosen the dust parameters, we can invert the observed
emission into a density profile. To do this, we start with a choice of
density distribution and predict its expected emission by solving the 
equation of radiative transfer and simulating an on-the-fly
observation with a bolometer array (including secondary chopping). The
result of this prediction is compared with the observed radial
profile, and the density distribution is corrected accordingly; the process 
is repeated 
until the model matches the observations. As density profiles, we have used
two families of curves. One is the empirical formula
$$n(r) = n_0/(1+(r/r_0)^\alpha),$$
where $n_0$, $r_0$, and $\alpha$ are free parameters that correspond to the 
central density, half maximum radius, and asymptotic power law of the
density profile. Note that this profile contains naturally the central
flattening first found by Ward-Thompson et al. (1994). The other family
of density profiles is the family of isothermal (Bonnor-Ebert) profiles, 
which has only two free parameters, the central density and the 
effective sound speed. 

The result from the fitting procedure is shown in Figure 3, where both 
linear-linear and log-log radial profiles of the observed 1.2mm intensity
(squares) are shown together with the predictions from the analytic
(solid lines) and isothermal (dashed-lines) models. As can be seen, both 
families of models fit similarly well the data, with the analytic model
for L1498 fitting slightly better the profile than the isothermal model. 
Both families of models require similar 
central densities, $10^5$ cm$^{-3}$ for L1498
and $2 \; 10^5$ cm$^{-3}$ for L1517B, and the size of the half maximum radius
(from the analytic model) is approximately 10,000 AU for L1498 and 
5,000 AU for L1517B. L1498, therefore, is slightly less dense and
more extended than L1517B, but both cores have similar central 
column densities: 3-4 $\times 10^{22}$ cm$^{-2}$ (30-40 Av). With 
respect to the $\alpha$ parameter,
we derive 3.5 and 2.5 for L1498 and L1517B, respectively. The 2.5 value for
L1517B makes the analytic fit indistinguishable from isothermal fit, as 
it can be shown that the analytic formula becomes an excellent approximation
to the Bonnor-Ebert sphere for the case of $\alpha=2.5$ (see Tafalla et al.
2004, Appendix B).

\subsection{Temperature}

To derive the gas temperature profiles of L1498 and L1517B, we use 
NH$_3$(1,1) and (2,2) observations. The combined analysis of these
two lines can provide a reliable and straightforward estimate of the gas 
kinetic temperature with minimal assumptions, and to achieve that,
we have updated the classical
NH$_3$ analysis of Walmsley \& Ungeretchs (1983) using a Monte Carlo
radiative transfer solution. The derived radial profiles of gas kinetic
temperature are shown in Figure 4 (left panel), and for both cores are 
relatively flat with a constant value of 10 K. 

\begin{figure}[h]
\begin{center}
\resizebox{12cm}{!}{\includegraphics{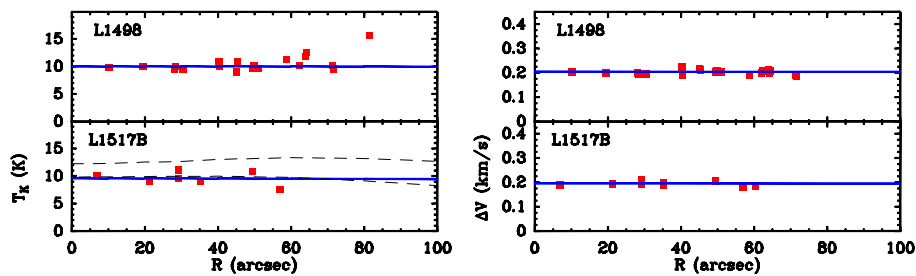}}
\end{center}
\caption{\protect\inx{\em Left: } radial profiles of gas kinetic temperature as
derived from the analysis of the NH$_3$ emission. The solid lines are
constant temperature fits at 10 K and the dashed lines are the predictions
from two models by Galli et al. (2002). {\em Right: } radial profiles of
total NH$_3$ linewidth as derived from a hyperfine analysis to the (1,1)
spectrum. The lines represent two constant linewidth fits with 
$\Delta$V = 0.2 km s$^{-1}$.}
\end{figure}

At the central gas densities of L1498 and L1517B, dust and gas
are expected to be thermally coupled, and therefore follow similar 
gradients with radius. As mentioned before, the dust kinetic temperature
is expected to drop towards the core center (Evans et al. 2001, Zucconi et 
al. 2001), but our gas temperature profiles do not show such a trend.
Our NH$_3$ observations have a $40''$ resolution (FWHM), so it is still
possible to hide a small temperature gradient, but not a very large 
one given the strong weight of the NH$_3$ emission towards the core 
center due to its higher abundance there (see below). To test our sensitivity 
to a temperature gradient, we have used the realistic temperature
profiles
predicted by Galli et al. (2002) for a core very similar to L1517B, and 
simulated an NH$_3$ observation with $40''$ resolution followed by 
a temperature analysis. The results, shown as dashed lines in Fig. 4,
indicate that to fit the data, a low cosmic rate is needed (1/5 of standard),
and that for this case, which has a central drop of about 1 K, the fit is
as good as the constant temperature fit. The central gas temperature in L1517B,
therefore, may be slightly lower than in the outside, but not by much more
than 1 K. Such a strong constrain arises from the strong sensitivity to
temperature of the NH$_3$(2,2), whose intensity would double by just 
increasing the temperature by 3 K.

\subsection{Turbulence}

The NH$_3$ lines also provide a sensitive measure of the gas turbulent motions
via the hyperfine analysis of the (1,1) spectra. Figure 5 (right panel) 
presents the radial
profiles of total  NH$_3$ linewidth (FWHM) for L1498 and L1517B, which are
again flat with a constant value of 0.2 km s$^{-1}$
and a very small dispersion ($\leq 0.01$ km s$^{-1}$,
consistent with random errors). This constant linewidth with radius 
is in contrast with the so-called linewidth-size
relation (e.g., Larson 1981) in two ways. First, the linewidth does not
increase with radius, and second, the linewidth of the more extended L1498
core is the same as that of the more compact L1517B. The lack of 
linewidth increase with radius in dense cores has been discussed
in detail by Goodman et al. (1998).

The total linewidth represented in Figure 4 is the harmonic average
of a thermal and a  non-thermal components. For NH$_3$ molecules at 
10 K, the thermal component is 0.16 km s$^{-1}$, so this 
component dominates the linewidth, while the turbulent 
component is subthermal (see also Myers 1983). To compare the contribution
of the two components to the equilibrium of the core, we take the ratio 
between thermal and non-thermal pressures:
$$ {P_{NT} \over P_T} = {\sigma_{NT}^2\over k T/m} \approx 0.07, $$
where $\sigma_{NT}$ is the turbulent component ($=\Delta V/\sqrt{8\; ln2}$
= 0.05 km s$^{-1}$)
and $m$ is the mean particle mass (2.33 the proton mass). 
As the ratio indicates,
the contribution of the turbulent component to the support of the core 
is negligible in the central 0.1 pc.

\subsection {Molecular composition}

Once the density, temperature and kinematics of the cores have been 
determined, the only free parameter we can adjust to model the observed 
emission of any molecule is the abundance profile of that species (see 
Tafalla et al. 2004 for a discussion on the velocity fields in L1498 
and L1517B, omitted here 
for lack of space). This means that we can use the L1498 and L1517B cores 
as laboratories of known
physical properties to measure the behavior of molecules 
at high densities and
determine their sensitivity to freeze out
onto dust grains. In order to simplify the modeling and the comparison
between molecules, we use for all 
species an abundance profile
with a constant value of $X_0$ at large radius and a step 
central hole at $R_{\rm hole}$ (NH$_3$ has a central increase). 
For each species at least two transitions have been observed, and for each
of them we model simultaneously both its radial profile of 
integrated intensity and 
its central emerging spectrum. This is done with a Monte Carlo radiative 
transfer code (Bernes 1979) together with a convolution routine
to simulate an astronomical observation. 

\begin{figure}[h]
\resizebox{\hsize}{!}{\includegraphics[angle=270,width=10cm]{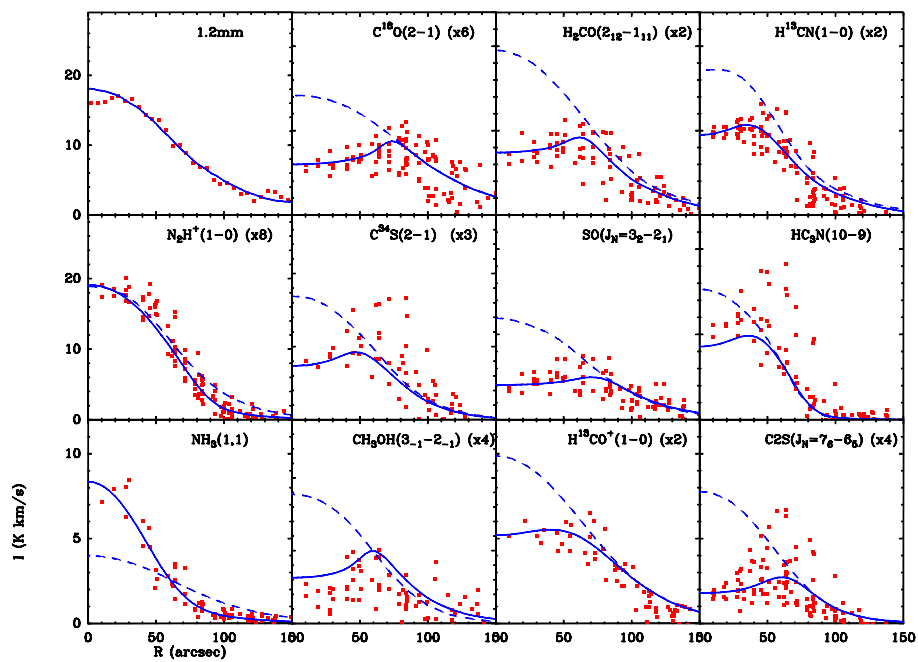}}
\caption{\protect\inx{Radial} profiles of emission from L1498 for the tracers
shown in Figure 2. The squares represent observations,
the dashed lines are the predictions from constant abundance models 
set to fit the outer emission, and the solid lines are the
best fit models (they also fit additional lines not shown here).
For all species in the second, third, and fourth columns, the best fit 
models contain a central abundance hole.}
\end{figure}

A summary of the modeling results is presented in 
Figure 5, where the radial profiles for one transition of each 
species in L1498 are shown (squares) together with the best fit 
model (solid line) and a model with constant abundance equal to
the outer value of the best fit model (dashed lines). As can be seen, all
molecules but N$_2$H$^+$ and NH$_3$ require a central abundance hole
of variable size (5,000-10,000 AU), which indicates that 
they freeze out at densities of a few 10$^4$ cm$^{-3}$. A similar behavior
is found in L1517B. 

The above pattern of abundances in L1498 and L1517B is in general agreement
with recent models of starless core chemistry, which show that only species
based on the simplest nitrogen chemistry can survive for a significant time at 
densities of 10$^5$ cm$^{-3}$ (e.g., Bergin \& Langer 1997, 
Shematovich et al. 2003, Aikawa et al. 2003). 
This behavior results from a combination of the lower binding energy 
of the N$_2$ molecule
to dust grains, which allows N chemistry
to proceed at high densities, together with the favorable chemistry
resulting from the disappearance of CO from the gas phase (see chapter
by Paola Caselli in this volume for more details). Although the
general agreement between models and observations is good, there 
are still some puzzling behaviors, like the abundance increase of NH$_3$
at high densities while the abundance of its cousin species N$_2$H$^+$ stays
constant. Further modeling of the high density core chemistry is still
needed to explain the data (see Aikawa et al. 2005).

From the point of view of observations, the results from L1498 and 
L1517B are a strong reminder
that very few molecules survive in the gas phase at the typical
densities of a starless core. Extreme care should therefore be 
exercised when choosing a molecular tracer to study the properties 
of the star-forming material in these environments.

\section{Tracing core evolution with molecular freeze out}

Our new understanding of molecular freeze out is
finally allowing a self consistent picture of core interiors to emerge. 
It also promises to shed crucial light on the 
still mysterious process of core contraction. This is so because molecular
freeze out is a progressive and irreversible process at the
densities and temperatures of starless cores. Each molecule that hits 
a dust grain under these conditions will stick to its surface and will not 
evaporate thermally (cosmic ray-induced evaporation seems not efficient
enough to reverse the process, see Leger et al. 1985, Hasegawa \& Herbst 1993).
The amount of freeze out in a given parcel of gas, therefore, will increase
with time, and it should be possible in principle to convert the
amount of freeze out into an estimate of the contraction age. In 
practice, unfortunately, this is not yet achievable because of the 
uncertainties in the binding energies of molecules to dust grains  
(but see Aikawa et al. 2005
for a recent detailed modeling). Molecular freeze out, however, can
already be used as a qualitative time marker thanks to the simple fact that
it increases with time. This means that by looking at the amount of
freeze out of a sensitive molecule (like CO or CS), we can distinguish 
cores at different contraction stages, although we cannot yet assign them
a particular age.

Molecular freeze out is not the only process that increases with time
in a contracting core. Some molecules like NH$_3$ and N$_2$H$^+$ 
gradually become more abundant because their
formation depends on the slow formation of N$_2$ (e.g., Suzuki et al. 1992),
so they are also useful indicators of core evolution 
(N$_2$H$^+$ is further enhanced by the depletion of CO).
Thus, we can expect that a core starts its life being CO/CS rich at its
center (not enough
time to freeze out) and NH$_3$/N$_2$H$^+$ poor (not enough time 
to form N$_2$) and evolves toward being centrally CO/CS poor
(due to freeze out) and NH$_3$/N$_2$H$^+$ rich (due to late-time
chemistry). To quantify this trend,
we define a ratio between the intensities of two molecules
with opposite behaviors. For observational convenience, we chose the
two thin species C$^{18}$O and N$_2$H$^+$, as they have J=1-0 transitions
at 3mm, and can therefore be observed with the same telescope
at similar resolutions. Thus, we define
$$R = I[{\rm C}^{18}{\rm O}(1-0)]\;/\;I[{\rm N}_2{\rm H}^+(1-0)],$$
where both integrated intensities are measured at the core center (as
defined by the mm continuum peak). A young core is expected to have
strong C$^{18}$O emission and weak N$_2$H$^+$ emission, so it will
be characterized be a relative large value of $R$. Conversely, an
evolved core will be weak in C$^{18}$O and bright in N$_2$H$^+$, so
its $R$ ratio will be low. Using typical abundances of C$^{18}$O and 
N$_2$H$^+$ in dense cores (Tafalla et al. 2004), we estimate that the
dividing line between young and evolved cores occurs at about $R=1$.

\begin{figure}[h]
\begin{center}
\resizebox{10cm}{!}{\includegraphics{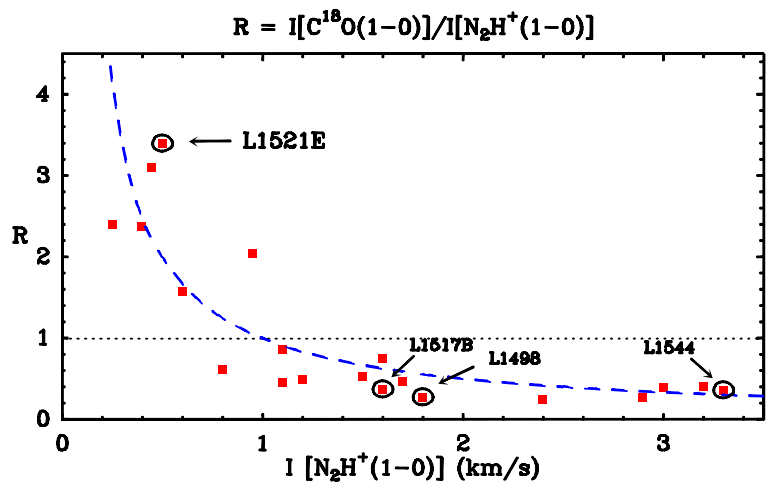}}
\end{center}
\caption{\protect\inx{Plot} of the $R$ factor 
($R = I[{\rm C}^{18}{\rm O}(1-0)]/I[{\rm N}_2{\rm H}^+(1-0)]$) 
as a function of N$_2$H$^+$(1--0) intensity for
a sample of starless cores. This factor seems to be an indicator of
core evolution, and to decrease as a core ages. The positions of the
extremely young core L1521E, the intermediate-age cores L1498 and L1517B,
and the highly evolved L1544 are indicated. The dashed line is the 
prediction from a simplified model of core evolution.} 
\end{figure}

Figure 6 presents a plot of the $R$ ratio as a function of N$_2$H$^+$(1--0)
central intensity for a sample of 21 starless cores.
Most cores lie below the $R=1$ line,
suggesting that 
C$^{18}$O is frozen out at their center, a fact confirmed
by Monte Carlo radiative transfer calculations for some of them 
(L1498, L1517B, L1544, see Tafalla et al. 2002).
As these cores were selected from NH$_3$ surveys
(like that of Benson \& Myers 1989), the presence of freeze out
at their center is an indication that they belong to a population 
of objects already chemically evolved, and therefore relatively
``old'' (of course, not old enough to have formed a star).

In order to find cores with a ratio $R>1$ indicative of
chemical youth, we have selected 
starless cores known to have weak
or undetected NH$_3$ emission, and we have mapped them in
C$^{18}$O and N$_2$H$^+$ with the 14m FCRAO telescope (Tafalla
et al. 2005, in preparation). 
As a result of this search, we have 
identified a small population of starless cores with  $R>1$,
which at the same time have a relatively low value of 
$I[{\rm N}_2{\rm H}^+(1-0)].$ These cores populate the upper-left 
region of the diagram in Fig. 6, and represent the best 
candidates for chemically young starless cores. The object
with largest $R$ value is L1521E in Taurus, and because of
its extreme characteristics, it has been the subject of 
a more detailed study.

\subsection{L1521E: the youngest starless core?}

To understand the origin of the extreme $R$ value 
in L1521E, we have observed this core 
with high resolution using
the IRAM 30m telescope (see Tafalla \& Santiago 2004
for a full report). These observations have been 
made in the 1.2mm continuum,  N$_2$H$^+$(1--0), and several 
rare isotopes of CO, and the results have been analyzed
with the same method as the L1498 and L1517B data presented 
in section 4. A gas and dust kinetic temperature of 10 K
has been assumed, as suggested by the observations of (very weak) 
NH$_3$ lines with the Effelsberg 100m telescope. 

\begin{figure}[h]
\resizebox{\hsize}{!}{\includegraphics{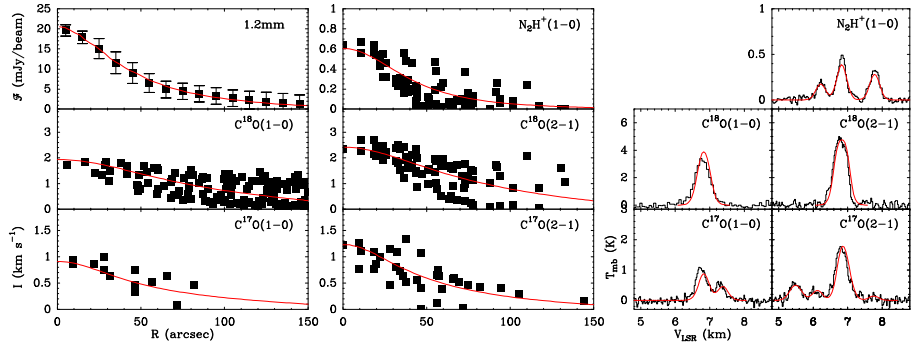}}
\caption{\protect\inx{Radial} profiles of integrated intensity (left)
and emerging spectra (right) of different tracers towards L1521E.
The squares and histograms are observed data, and the lines are
model predictions. In contrast with the results from the L1498 
analysis (Figure 5), the CO isotopomer
emission from L1521E can be fit with a constant abundance
model. This makes L1521E the first starless cores with negligible
CO depletion.}
\end{figure}

The main results of the L1521E analysis are summarized in
Fig. 7, which shows radial profiles of integrated 
intensity and emerging central spectra for all our data. As in the
L1498 and L1517B analysis, the squares represent the
observed data and the lines are fits from a Monte Carlo
radiative transfer calculation (or an optically thin 
analysis for the 1.2mm continuum). In contrast with the
L1498 and L1517B cores, however, all 
CO isotopomer data from L1521E can be fit assuming a constant 
abundance, together with a C$^{18}$O/C$^{17}$O isotopic ratio of 3.65
(Penzias 1981). In addition, the 
value of the C$^{18}$O abundance in L1521E is 
almost the same as the value needed to fit the outer part of
the L1498 and L1517B cores. This all suggests that C$^{18}$O
is not frozen out at the center of L1521E, or if it is, it
happens only in a very small region diluted by our 
resolution and sampling (about $20''$). In this respect, 
L1521E is the first starless core consistent with no freeze out.

The very large $R$ value of L1521E ($R=3.4$) not only
arises from a lack of significant C$^{18}$O depletion. It
also results from a very low value of its N$_2$H$^+$ abundance, which
according to the Monte Carlo calculations is 8 times lower than
that estimated for L1498 and L1517B. As mentioned before, a
low  N$_2$H$^+$ abundance is another indication of extreme youth
given the late-time character of this species. More evidence
for a L1521E being a young core comes from
previous work by Suzuki et al. (1992) and 
Hirota et al. (2002), who also concluded 
that L1521E is unusually young based in its very bright CS and
C$_2$S emission and its weak NH$_3$ lines. It is reassuring that 
all these different lines of work consistently indicate 
that the L1521E core is unusually chemically unprocessed.

Given the multiple indications of chemical youth, it is 
surprising that the density profile of L1521E is very
similar to that of more evolved cores like L1517B. At
least naively, one would have expected that a young core should
have a lower central density and a more distended (cloud-like) structure,
but clearly this is not the case for L1521E. Whether this is
unique to this core (as a the result of fast contraction)
or typical of all  young cores requires the study of
more objects that lie above the $R=1$ line in Figure 6. Work
currently in progress will hopefully answer this question
soon.

%

\begin{acknowledgments}
It is a pleasure to thank the organizers of this workshop, 
especially Nanda Kumar, for an enjoyable and productive
meeting. Part of the work presented here is the result of an 
ongoing collaboration with Joaquin Santiago, Phil Myers,
Paola Caselli, Malcolm Walmsley, and Claudia Comito. I thank them warmly 
for help and many discussions on starless cores and related issues 
over the last several years.
\end{acknowledgments}

\end{document}